\documentstyle[aps,prb,floats,twocolumn]{revtex}
\input epsf     

\begin{document}
\draft
\title{Point-charge electrostatics in disordered alloys}
\author{C. Wolverton, Alex Zunger, S. Froyen, and S. -H. Wei}
\address{National Renewable Energy Laboratory, Golden, CO 80401\\}
\date{May 21, 1996}
\maketitle
%
%
%
{\let\clearpage\relax
\twocolumn[%
\widetext\leftskip=0.10753\textwidth \rightskip\leftskip
\vspace{-11pt}
\begin{abstract}
A simple analytic model of point-ion electrostatics 
has been previously proposed
(R. Magri, S. -H. Wei, and A. Zunger, 
Phys.\ Rev.\ B {\bf 42}, 11388 (1990))
in which the magnitude of the 
net charge $q_i$ on each atom in an ordered
or random alloy depends linearly on the number $N_i^{(1)}$ of
unlike neighbors in its first coordination shell.
Point charges extracted from recent 
large supercell (256-432 atom) local
density approximation (LDA) 
calculations of Cu$_{1-x}$Zn$_x$ random alloys
now enable an assessment of the physical
validity and accuracy of the
simple model.  We find that this model accurately
describes
(i) the trends in $q_i$ {\em vs.} $N_i^{(1)}$, particularly
for fcc alloys,
(ii) the magnitudes of total electrostatic energies
in random alloys,
(iii) the relationships between constant-occupation-averaged
charges $\langle q_i \rangle$ and Coulomb
shifts $\langle V_i \rangle$ 
(i.e., the average over all sites occupied by either $A$ or $B$ atoms)
in the random alloy, and
(iv) the linear relation between the site charge $q_i$
and the constant-charge-averaged Coulomb shift $\overline{V}_i$ 
(i.e., the average over all sites with the same charge) 
for fcc alloys.
However, for bcc alloys the {\em fluctuations} predicted 
by the model in the $q_i$ {\em vs} $V_i$
relation exceed those found in the LDA supercell
calculations.
We find that 
(a) the fluctuations present in
the model have a vanishing contribution
to the electrostatic energy.  
(b) Generalizing the
model to include a dependence of the charge on
the atoms in the first {\em three (two) shells}
in bcc (fcc) - rather than the first shell only - 
removes the fluctuations,
in complete agreement with the LDA data.
We also demonstrate an efficient way to extract
charge transfer parameters of the generalized model from
LDA calculations on small unit cells.
\end{abstract}
\vspace{11pt}
\pacs{PACS numbers: 61.66.Dk, 71.10.+x, 61.50.Lt}
]}
\narrowtext

\section{Introduction}

The structural stability of alloys and compounds
is determined by the kinetic, electrostatic, and 
exchange-correlation contributions to the total energy.
In first-principles
calculations based on Hartree-Fock or on density functional theory, 
the electrostatic portion of the total energy is characterized
in terms of the electronic charge density $\rho({\bf r})$ and
the nuclear charges $z_i$.  For systems 
with uniquely specified nuclear positions $\{{\bf R}_i\}$ and
charges $\{z_i\}$, the
charge density is a well defined quantity
as is the electrostatic (el) portion of the total energy:
\begin{eqnarray}
\label{e.lda}
E_{\rm el} &=& \frac{1}{2} \; \int \int d^3{\bf r} \; d^3{\bf r'} \;
\frac{\rho({\bf r})\rho({\bf r'})}
{| {\bf r} - {\bf r'} |}
- \sum_i \int d^3{\bf r} \;
\frac{\rho({\bf r})z_i}
{| {\bf r} - {\bf R}_i |} \nonumber \\
&+& \frac{1}{2} \sum_i \sum_j 
\frac{z_iz_j}{| {\bf R}_i - {\bf R}_j |}
\end{eqnarray}
Indeed, in many previous calculations on 
ordered structures 
\cite{Zunger94,Wei90,Sluiter90,Sanchez91,Lu91,Asta93,Lu94,Wolverton95a} and
``supercell'' models of random alloys, \cite{SQS,Lu91a} 
there are
well-defined $\{{\bf R}_i ; z_i\}$, so the
electrostatic energy was obtained from Eq. (\ref{e.lda}).
However,
in simpler approaches,\cite{Magri90,Borici93,Johnson93,Wolverton95,Ruban95} 
one approximates 
the electrostatic energy
by replacing the continuous charge density $\rho({\bf r})$, 
with fictitious point charges $q_i$ at each site $i$.   
For a system with $N$ sites,
the electrostatic or Madelung (M) energy is 
\begin{equation}
\label{e.point}
E_{\rm M} = \frac{1}{2N} \sum_i  \sum_{j \neq i} \frac{q_iq_j}{R_{ij}}
\end{equation}
where $R_{ij}$ is the distance between sites $i$ and $j$.
The Madelung energy may also be written
\begin{equation}
\label{e.point.2}
E_{\rm M} = \frac{1}{2N} \sum_i q_i V_i
\end{equation}
where $V_i$ is the Coulomb shift at site $i$ due to all charges
other than $q_i$:
\begin{equation}
\label{shift}
V_i = \sum_{j \neq i} \frac{q_j}{R_{ij}}.
\end{equation}
The point charges are obtained 
by partitioning $\rho({\bf r})$ into ``domains''
(spheres, polyhedra, etc.) and integrating the total charge
in each domain.  However, because there is not a unique
way to partition a three-dimensional space, the
point charges are not uniquely defined.

For periodic systems 
(e.g., ordered structures with a {\em primitive}
cell or random structures defined by 
{\em supercells} 
\cite{Lu91,SQS,Lu91a,Magri90,Wang95,Faulkner95,Faulkner.conf})
where all sites $i$ are defined as distinct 
entities (not as averages) and $q_i$ and ${\bf R}_i$ are
specified, $E_M$ can be readily computed from Eq. (\ref{e.point})
using, for example, the Ewald method.
In most statistical approaches to alloys (e.g., the coherent
potential approximation, or CPA) \cite{cpa} however,
one attempts a description of a random alloy without a
specification of all {\em distinct} sites $i$ but rather
some averages over $i$.  In such approaches one calculates
the Madelung energy of the random alloy by determining
the configurationally averaged correlation between charges
$\langle q_i q_j \rangle$, and using
\begin{equation}
\label{e.point.cpa}
\langle E_{\rm M}\rangle_R = \frac{1}{2N} \sum_i \sum_{j \neq i}
\frac{\langle q_iq_j\rangle}{R_{ij}}.
\end{equation}
Until 1990, all CPA-based 
models for alloy energies have assumed uncorrelated charges
\begin{equation}
\label{charge.uncorr}
\langle q_iq_j\rangle = \langle q_i\rangle \langle q_j\rangle
\end{equation}
which leads to a vanishing electrostatic energy for the 
random alloy
\begin{equation}
\label{e.point.cpa.zero}
\langle E_{\rm M} \rangle_R = 0
\end{equation}
on account of electroneutrality.  This approximation 
[Eq. (\ref{charge.uncorr})] was based on the 
expectation that a random (i.e. uncorrelated)
distribution of {\em atoms} on sites would lead to
an equally random distribution of {\em charges},
i.e., the charge on an atom in a given alloy
is a property of the atom, irrespective of its environment.
Eq. (\ref{e.point.cpa.zero}) has been assumed in
many CPA-based calculations 
\cite{cpa,Johnson90,Pinski91,Turchi91,Johnson94,Weinberger94} 
involving the total energy of random alloys.
Magri {\em et al.} \cite{Magri90} subsequently criticized this
approach as being physically implausible, since the
assumption of uncorrelated charges
[Eq. (\ref{charge.uncorr})] means that an $A$ atom
surrounded locally by only $A$ atoms will have the same
charge as an $A$ atom surrounded by $B$ atoms;  chemical intuition
suggests, however, that the charge on a site will depend on
the identity of atoms in its environment because charge transfer
is present only between {\em dissimilar} sites.

Magri {\em et al.} \cite{Magri90} 
noted that in a random alloy, even though 
the {\em occupation} of site $i$ is independent of the occupation
of other sites by definition, 
the {\em charges} on a site do depend on the
occupations of other sites.
These authors therefore
proposed a simple model
to describe the magnitude of point charges in disordered (and
ordered) alloys:  The magnitude of the charge on a site is
linearly proportional to the number $N_i^{(1)}$
of unlike nearest neighbors surrounding that site.
With this charge model, Magri {\em et al}.\
went on to demonstrate that even for the case of a random alloy
with completely uncorrelated atomic occupations, 
charge correlations exist in the alloy and these correlations 
lead to a non-zero Madelung energy.  

Subsequent to the proposal of Magri {\em et al.}, \cite{Magri90}
the charge model has been used in many contexts:

(i) Lu {\em et al.} \cite{Lu91a} showed that LDA calculations on
ordered compounds produced charge densities which, when integrated inside
muffin-tin spheres,
gave point charges which reproduced the behavior of 
the model.
They also examined \cite{Lu91} the effect of the ensuing 
electrostatic energy of the random alloy on
the sign of the ordering energy.

(ii) Abrikosov {\em et al.} \cite{Abrikosov92} and 
Johnson and Pinski \cite{Johnson93} derived corrections
to the CPA total energy 
which introduced charge correlations in random alloys.
These corrections were shown to be 
consistent with the charge model of Magri {\em et al.}
Several authors subsequently
used these corrections in total-energy CPA calculations
to determine lattice constants and formation energies of
random metallic alloys, finding significant effects due to 
charge correlations:  Johnson and Pinski \cite{Johnson93}
estimated the total energy contribution due to charge correlations to
be $-1.25$, $-5.3$, and $-7.7$ mRy/atom for Cu$_{0.5}$Zn$_{0.5}$,
Cu$_{0.5}$Au$_{0.5}$ and Ni$_{0.5}$Al$_{0.5}$ alloys, respectively.
(Typical values of alloy formation energies are 
$\sim$10-20 mRy/atom.)
Korzhavyi {\em et al}.\ found \cite{Korzhavyi94}
that the energetic contribution due to 
charge correlations for Al$_{0.5}$Li$_{0.5}$ is 
-16.0 mRy/atom which results in a change of {\em sign} in the 
formation energy of Al-Li alloys.

(iii) Borici and Monnier \cite{Borici93} used the charge model to
study the segregation behavior of a semi-infinite random
Madelung lattice.  
For semi-infinite surface geometries,
these authors found that charge correlations 
lead to monotonic surface segregation
profiles and 
a segregation of the minority species to the surface.
On the other hand, for thin-film geometries
charge correlations lead to
oscillatory surface segregation profile, and an
enrichment of the majority species on the surface.

(iv) Wolverton and Zunger \cite{Wolverton95} determined 
the ground state long-range order and the high-temperature 
short-range order of fcc-, bcc-, and sc-based alloys 
due to electrostatic effects.
These authors also showed \cite{Zunger94,Wolverton95}
how the charge model could be analytically
mapped onto a cluster expansion, which allowed for the efficient
and accurate determination of energies of any ordered or disordered
configuration without the use of Ewald methods.

(v)  Ruban {\em et al.} \cite{Ruban95} compared the energies 
of charge-correlated
CPA calculations with ordered compound LDA calculations to 
determine the optimum prefactor for the electrostatic energy
for Cu-Au and Ni-Pt alloys.  The energetic contribution
due to charge correlations was again found to be significant:
For instance, for random Cu$_{75}$Au$_{25}$ alloys, 
electrostatic
contributions to the total energy were found to lower the
mixing energy by a factor of $\sim$3-6 
relative to CPA calculations
with a complete neglect of charge transfer effects.

The charge model ansatz of Magri {\em et al.}
was thus far tested by comparing
its charges $\{q_i\}$ with those found
in small-unit-cell
($\leq$16 atom) LDA calculations, and also for only one
lattice type - fcc.  Recently, much larger
LDA supercell calculations became available \cite{Wang95} 
for fcc {\em and} bcc-based alloys.
These calculations
combine a locally self-consistent muffin-tin scheme
with a massively parallel computer enabling 
LDA calculations on 256- and 432-atom supercells for 
random Cu-Zn alloys.  
\cite{Faulkner95,Faulkner.conf} 
Faulkner {\em et al}.\ \cite{Faulkner95,Faulkner.conf}
have used the charge density from these large LDA supercells to 
examine the behavior of point charges $\{q_i\}$
in random Cu-Zn alloys, finding interesting relations
between charges and certain potentials. 
Here we determine
to what extent the simple charge model is able
to describe the electrostatic properties
of complicated large scale (256-432 atom) LDA based calculations.
We find that the model works very well for fcc lattices,
but that in bcc lattices, where the first few coordination shells
are near to one another, the charge on a site is correlated
with the occupations on {\em a few} neighboring shells,
not just one.  The effects of such corrections to the
total electrostatic energy $\langle E_M \rangle_R$ are
small, however.

\section{The Simple Charge Model}

Consider an $A_{1-x}B_x$ alloy with $N$ sites and a nearest
neighbor coordination
number $Z$.
The model of Magri {\em et al.} \cite{Magri90} is based on 
the assumption that the excess charge on a site depends only on
the identity of its {\em first} neighbors.  If an
$A$ atom on a central site 
is surrounded purely by $Z$ atoms of type $A$,
the charge is taken to be zero.  
If it is surrounded by $Z$ atoms
of type $B$, the charge is maximal, $2Z\lambda$.  For
intermediate occupations of the first coordination
shell, we assume a linear interpolation between these
two limits.  Formally, we then write this charge as:
\begin{equation}
\label{charge}
q_i = \lambda \sum_{k=1}^{Z} [ \hat{S}_i - \hat{S}_{i+k} ],
\end{equation}
where the pseudospin $\hat{S}_i$ is
-1 (+1) if an $A(B)$ atom is located at site $i$.  (The set of
variables $\hat{S}_i$ for all sites $i$ defines the configuration
$\sigma$.)  $\hat{S}_{i+k}$
indicates the occupation of the $Z$ lattice sites which are
nearest neighbors to $i$, and hence the summation in Eq. (\ref{charge})
indicates the number of unlike nearest neighbors surrounding the
site $i$. $\lambda$ is a constant which
indicates the magnitude of the charge transfer and
is an undetermined parameter of the 
model.  Thus, the charge model will give trends
in the behavior of physical properties, but will not give
numerical values of properties without some input value of
$\lambda$.

Several questions may be asked concerning the
parameter $\lambda$:
1) Should $\lambda$ be explicitly composition-dependent?
2) Should $\lambda$ be explicitly volume-dependent?
Since the equilibrium volume is a function of composition
in size-mismatched alloys, an explicit volume dependence of
$\lambda$ would lead to an implicit composition-dependence.
It is important to physically 
distinguish between these two dependences.
3) Should the values of $\lambda$ be extracted from large unit cell 
or small unit-cell alloys i.e., does $\lambda$ contain mostly
short-range or long-range information?
Values of $\lambda$ have been estimated by LDA calculations,
\cite{Lu91a,Johnson93,Faulkner.conf}
ranging from small unit cell ordered compounds ($\sim$8-16 atoms), 
up to large LDA simulations of random alloys ($\sim$200-400 atoms).  
For computational simplicity, one should know whether it is 
equally valid to extract
values of $\lambda$ from ordered or random alloys, and whether
one can even use smaller cells ($\sim$2-4 atoms) than have been
currently used.

We next examine the physical consequences of charges which obey
Eq. (\ref{charge}). We then compare these consequences 
with results of LDA supercell calculations in order to assess the
physical validity of the model. 
With regard to the questions raised above, we demonstrate that
the simple charge model represents well the charge transfer of {\em different}
unrelaxed configurations at a {\em common} volume.
If more than one volume is considered (e.g., for a lattice-mismatched
alloy at more than one composition), the parameter of the model $\lambda$
would presumably need to be explicitly volume-dependent (implicitly
composition-dependent).  Also, we find that values of $\lambda$
extracted from 2-4 atom LDA calculations agree favorably with those
extracted from much larger 200-400 atom LDA calculations, thereby 
resulting in a drastic computational simplification.

\section{Physical Consequences of the Charge Model}

\subsection{Average Charges}

The average charge {\em on all sites}, $\langle q \rangle$ 
is defined as:
\begin{equation}
\langle q \rangle = \frac{1}{N} \sum_i q_i
\end{equation}
Combining this with Eq. (\ref{charge}) gives
$\langle q \rangle  =  0$,
as guaranteed by global charge neutrality.
However, what is more interesting is the constant-occupation-average
$\langle q \rangle_A$ (or $\langle q \rangle_B$), i.e., 
the average
charge of all sites occupied by A (B) atoms.  This 
constant-occupation-average is a function of the 
configuration $\sigma$ and composition $x$,
and we can analytically derive this quantity for any
arbitrary configuration.  The definition of
$\langle q \rangle_A$ is
\begin{equation}
\langle q \rangle_A = \frac{1}{N_A} \sum_i q_i \; \Gamma_i^A \; ,
\end{equation}
where $N_A$ is the number of $A$ atoms in
$\sigma$ and $\Gamma_i^A$ is the Flinn operator such that
$\Gamma_i^A$ = 1 if site $i$ is occupied by an $A$ atom,
and $\Gamma_i^A$ = 0 otherwise.  The Flinn operator
is given by
$\Gamma_i^A$ = $(1-\hat{S}_i)/2$.
Thus,
\begin{eqnarray}
\label{chargea}
\langle q \rangle_A &=& \frac{\lambda}{2N_A}
\sum_i \sum_k (\hat{S}_i - \hat{S}_{i+k} - \hat{S}_i^2 
+ \hat{S}_i\hat{S}_{i+k})
\nonumber \\
\mbox{} &=& -\frac{Z\lambda}{2(1-x)} (1 - \overline{\Pi})
\end{eqnarray}
where $\overline{\Pi}$ is the nearest neighbor (NN)
pair correlation function, i.e., 
the lattice average of $\Pi_{i,j} = \hat{S}_i\hat{S}_j$ 
for $i$ and $j$ NN.
A similar analysis gives
\begin{equation}
\label{chargeb}
\langle q \rangle_B = \frac{Z\lambda}{2x} (1 - \overline{\Pi})
\end{equation}
In addition, the difference $\Delta$ in 
constant-occupation-averaged charges is given by
\begin{equation}
\label{delta}
\Delta = \langle q \rangle_B - \langle q \rangle_A = 
2Z\lambda \frac{1 - \overline{\Pi}}{1-\langle S \rangle^2}
\end{equation}
where $\langle S \rangle = 2x-1$.
Equations (\ref{chargea}), (\ref{chargeb}), and (\ref{delta}) 
are quite general and apply to any configuration (ordered,
random, or partially ordered).  These expressions may be 
evaluated in various classes of configurations which are interesting:

{\em Random Alloys}:
We define a random alloy as one in which the 
{\em occupation variables} $\hat{S}_i$ are
uncorrelated, i.e.
$\overline{\Pi}_{i,j} = \langle S_i \rangle \langle S_j \rangle
\equiv (2x-1)^2$.
As we will see below, this does not imply that the {\em charges}
$q_i$ are uncorrelated.
In a random alloy, we have
\begin{equation}
\label{charge.rand}
\langle q \rangle_A = -2Z\lambda x \; \;  ;  \; \;
\langle q \rangle_B = 2Z\lambda(1-x) \; \;  ;  \; \;
\Delta = 2Z\lambda
\end{equation}
Interestingly, $\Delta$ for random alloys is independent of
alloy composition $x$.

{\em Short-range Ordered Alloys}:
Short-range order (SRO) measures the extent to which there
are atom-atom correlations in a disordered alloy.
The relation
between the nearest neighbor
Warren-Cowley SRO
parameter ($\alpha$) and $\overline{\Pi}$ is
\begin{equation}
\alpha = \frac{\overline{\Pi} - \langle S \rangle^2}
{1 - \langle S \rangle^2}.
\end{equation}
This expression combined with Eqs. (\ref{chargea}) and
(\ref{chargeb}) and (\ref{delta}) gives
the constant-occupation-averaged charges 
for alloys possessing some degree of SRO:
\begin{eqnarray}
\label{charge.sro}
\langle q \rangle_A = -2Z\lambda x (1-\alpha)\; \;  &;&  \; \;
\langle q \rangle_B = 2Z\lambda(1-x) (1-\alpha) \; ; \nonumber \\
\Delta &=& 2Z\lambda(1-\alpha)
\end{eqnarray}
So, the difference in charges $\Delta$ should increase
in an ordering type alloy ($\alpha < 0$) relative to the
random values, but should decrease in a clustering type
alloy ($\alpha > 0$).

{\em Long-range Ordered Alloys}:
Long-range order (LRO) gives an indication for the relative
population of $A$ or $B$ atoms on a given sublattice.
The extent of LRO in an alloy may be 
described by (one or more) LRO parameters $\eta$.
For example, for an alloy at $x$=1/2 with a single 
LRO parameter $0 \le \eta \le 1$ 
(and no correlations between atoms on the same sublattice),
$\overline{\Pi}(\eta) = \eta^2\overline{\Pi}(1)$,
so for any state of LRO at $x$=1/2:
\begin{eqnarray}
\label{charge.lro}
&& \langle q \rangle_A(\eta) = 
-Z\lambda [1-\eta^2\overline{\Pi}(1)]\; \;  ;  \; \; \nonumber \\
&& \langle q \rangle_B(\eta) =  
 Z\lambda [1-\eta^2\overline{\Pi}(1)]\; \;  ;  \nonumber \\
&& \Delta(\eta) = 2Z\lambda[1-\eta^2\overline{\Pi}(1)]
\end{eqnarray}
For example, in CsCl ($B2$) type ordering
$\overline{\Pi}(1) = -1$, thus
as the degree of LRO increases, $\Delta$ increases due to
the increased number of unlike nearest neighbors.

\subsection{Constant-Occupation-Averaged Coulomb Shifts, 
$\langle V \rangle_A$ and $\langle V \rangle_B$}

The Coulomb shift $V_i$ [Eq. (\ref{shift})] averaged
over all sites $\langle V \rangle$
is zero (just as $\langle q \rangle$ is zero) due to global
neutrality.
A more interesting quantity is the 
constant-occupation-average of
the Coulomb shifts
on all sites occupied by A-atoms in a random alloy:
\begin{equation}
\langle V \rangle_A = \frac{1}{N_A} \sum_i \langle V_i \; \Gamma_i^A \rangle.
\end{equation}
where the latter brackets denote a configurational average.
This expression can be evaluated to give:
\begin{eqnarray}
\label{shifta}
\langle V \rangle_A &=& \frac{\lambda}{2(1-x)}
\langle \sum_{j \neq i} \sum_k \frac{1}{R_{ij}} 
(\hat{S}_j - \hat{S}_{j+k})(1 - \hat{S}_i) \rangle
\nonumber \\
\mbox{} &=& \frac{\lambda}{2(1-x)}
\sum_{j \neq i} \sum_k \frac{1}{R_{ij}} [1 - (2x-1)^2 ] \delta_{i,j+k}
\nonumber \\
&=& \frac{2xZ\lambda}{R_1}
\end{eqnarray}
where $R_1$ is the nearest-neighbor distance and
in the second equality of Eq. (\ref{shifta}), we have used
the orthonormal properties of the 
products of $\hat{S}_i$. \cite{SDG}
Similarly for B-atoms,
\begin{equation}
\langle V \rangle_B = -\frac{2(1-x)Z\lambda}{R_1}
\end{equation}

\subsection{Relation between Constant-Occupation-Averaged 
Charge and Coulomb Shift}

{}From Eqs. (\ref{charge.rand}) and (\ref{shifta}), we have the following
relation between constant-occupation-averaged charge 
and Coulomb shift, as predicted by the charge model:
\begin{equation}
\label{qave.Vave}
\langle V \rangle_{A,B} = \frac{-\langle q \rangle_{A,B}}{R_1}
\end{equation}

\subsection{Charge-Charge Correlation Functions}

The charge-charge correlation function between sites $i$ and $j$
is given by:

\begin{eqnarray}
\label{corr}
\langle q_iq_j \rangle &=& \frac{1}{N} \sum_{m=1}^N
q_{i+m}q_{j+m} \nonumber \\
\mbox{} &=& \lambda^2
[ Z^2 \overline{\Pi}_{i,j} - Z (\sum_{k=1}^Z \overline{\Pi}_{i+k,j}
+ \sum_{k'=1}^Z \overline{\Pi}_{i,j+k'} )  \nonumber \\
&+& \sum_{k=1}^Z \sum_{k'=1}^Z \overline{\Pi}_{i+k,j+k'} ]
\end{eqnarray}
The sums over $k$ ($k'$) are
over the nearest neighbors of site $i$ ($j$).
Eq. (\ref{corr}) is generally valid for any configuration and
any composition.  
For random alloys (i.e., alloys with uncorrelated site
occupations), the charge correlations for the $m$th shell
$\langle q_0q_m \rangle$ have been previously
derived \cite{Lu91} and are given by:
\begin{eqnarray}
\label{corr.random}
\langle q_0q_0 \rangle &=& 4x(1-x) \lambda^2 (Z_1^2 + Z_1) \nonumber \\
\langle q_0q_1 \rangle &=& 4x(1-x) \lambda^2 (-2Z_1 + K_1) \nonumber \\
\langle q_0q_m \rangle &=& 4x(1-x) \lambda^2 (K_m) \; \; ; \; \; m > 1
\end{eqnarray}
In these expressions, $Z_m$ is the coordination of the $m$th shell
(i.e., $Z_1 \equiv Z$),
and $K_m$ is the number of nearest-neighbor atoms shared
by sites $i$ and $i+m$.  
As found by Magri {\em et al.} \cite{Magri90},
Eqs. (\ref{corr.random}) demonstrate that even when the
{\em occupations} of sites are uncorrelated, the {\em charges} 
on these sites, obeying Eq. (\ref{charge}), are correlated.

\subsection{Electrostatic Energies of Random Alloys}

Using the charge-charge correlations in Eq. (\ref{corr.random}), 
one can obtain the electrostatic energies of random alloys 
which are a consequence of the charge model.
These energies of random alloys have been derived 
previously for fcc, bcc, and sc-based alloys. \cite{Magri90,Wolverton95}
The energies of fcc and bcc-based random alloys are given
by
\begin{eqnarray}
\label{e.rand}
\langle E_M \rangle_R^{\rm fcc}/E_0  &=&  -4x(1-x) 0.7395181... \nonumber \\
\langle E_M \rangle_R^{\rm bcc}/E_0  &=&  -4x(1-x) 0.3457752... 
\end{eqnarray}
where $E_0 = (16\lambda)^2/2R_1$.

\subsection{The $q-V$ Relation between Charges and Coulomb Shifts}

For a completely random alloy, we may analytically
derive (Appendix A)
a relation between the charges $q_i$ and Coulomb
shifts $V_i$ from the charge model:
In the model of Eq. (\ref{charge}),
the magnitude of the 
charge on a site $i$ surrounded by $N_i^{(1)}$ unlike 
neighbors in the first shell does not depend on
the spatial configuration of the $N_i^{(1)}$ and is 
$| \, q_i[N_i^{(1)}] \, | = 2\lambda N_i^{(1)}$.  The Coulomb shift 
$V_i[N_i^{(1)}]$ on a site
surrounded by $N_i^{(1)}$ unlike nearest neighbors 
on the other hand, does depend in the model of Eq. (\ref{charge})
on the spatial configuration of the $N_i^{(1)}$ unlike atoms
around $i$ and also depends on more distant neighbors.  
If we {\em average} over all sites
having $N_i^{(1)}$ unlike neighbors,
we find analytically 
(Appendix A) the linear relation between charge $q_i[N_i^{(1)}]$ and
the constant-charge-averaged 
Coulomb shift $\overline{V}_i[N_i^{(1)}]$:
(i.e., an average over all sites with the same charge, and
hence with the same $N_i^{(1)}$), 
\begin{equation}
\label{qV.analytic}
q_i[N_i^{(1)}] \propto \overline{V}_i[N_i^{(1)}], 
\end{equation}
where $\gamma$ (in Ry$^{-1}$) is the
slope of this linear relation:  
\begin{eqnarray}
\label{gamma}
\gamma_{\rm fcc}(x=1/2) &=& 0.132R_1 \nonumber \\
\gamma_{\rm bcc}(x=1/2) &=& 0.163R_1
\end{eqnarray}
Note that $\overline{V}_i$ is a constant-charge-average
(still leaving the $N_i^{(1)}$ dependence), in contrast
to the constant-occupation-averaged $\langle V \rangle_A$.
To evaluate the {\em fluctuations} in $V_i(N_i^{(1)})$
about $\overline{V}_i$, we 
perform large-unit-cell simulations 
for a single, randomly selected configuration.  Equal numbers of
atomic types $A$ and $B$ ($x$=1/2)
are distributed at random over the
256 fcc sites and 432 bcc sites of the simulation cell.  
Point charges $\{q_i\}$ are then assigned by the 
model of Eq. (\ref{charge}).
Using the Ewald method, we then calculate the Coulomb shifts $V_i$ 
[Eq. (\ref{shift})] for each site in the cell.  
This gives a $q_i$ {\em vs.} $V_i$ relation for the charge model.

\section{Comparison:  Simple Model vs. LDA Simulations}

Recently, \cite{Faulkner95,Faulkner.conf} large scale
LDA supercell
calculations (256- and 432-atom) have been performed
for Cu-Zn alloys with Cu and Zn atoms placed randomly
on the fcc or bcc lattice sites of the supercell.
These calculations utilize a multiple scattering
framework, and are locally self-consistent:
The charge density associated with each atom 
is constructed
by considering only the electronic multiple scattering
processes in a finite spatial region (several neighboring
shells) centered at that atom.  
These LDA calculations also use the muffin-tin approximation:
The charge within the Wigner-Seitz cell surrounding each
site $i$ (volume $\Omega_i$) is made of a spherically
symmetric portion $\rho^i({\bf r}) = \rho_{\rm MT}^i(r)$
inside each muffin-tin sphere ($r < R_{\rm MT}^i$) and
is equal to a constant $\rho_0$ in the interstitial region
between spheres.
Point charges are then extracted from the muffin-tin charge
density by performing the following integral:
\begin{equation}
q_i = 4\pi \int_0^{R_{\rm MT}} \rho_{\rm MT}^i (r) r^2 dr
+ \rho_0 [ \Omega_i - \frac{4\pi}{3} (R_{\rm MT}^i)^3 ]
- z_i
\end{equation}
The calculations are carried out for unrelaxed geometries, thus
each of the 256 or 432 atoms has an equivalent Wigner-Seitz cell.
Cu and Zn have a 
very small electronegativity difference, so Madelung energy is
quite small in Cu-Zn alloys, and could be more susceptible to 
any errors in the calculation.  Cu-Zn is therefore
a critical test of any charge model, as the electrostatic effects 
in this system are quite subtle.
An alloy system with more robust charge transfer
(larger electronegativity difference)
could therefore be of interest in comparing magnitudes of 
electrostatic energies, charges, and other properties.

Differences of $\sim 2\%$ are cited \cite{Faulkner95}
between LDA calculations using bcc cells of 256 and 432 sites, 
and hence indicate a typical error 
due to using a finite-sized supercell.  An additional
consideration is that one, single configuration is considered
rather than a configurational average.  Thus,
in comparing properties of the charge model with
those of LDA supercell calculations, only disparities of more
than a few percent should be considered meaningful.

\subsection{Dependence of Charges on the Nearest-neighbor Environment}

%
%
\begin{figure}[tb]
\hbox to \hsize{\epsfxsize=0.80\hsize\hfil\epsfbox{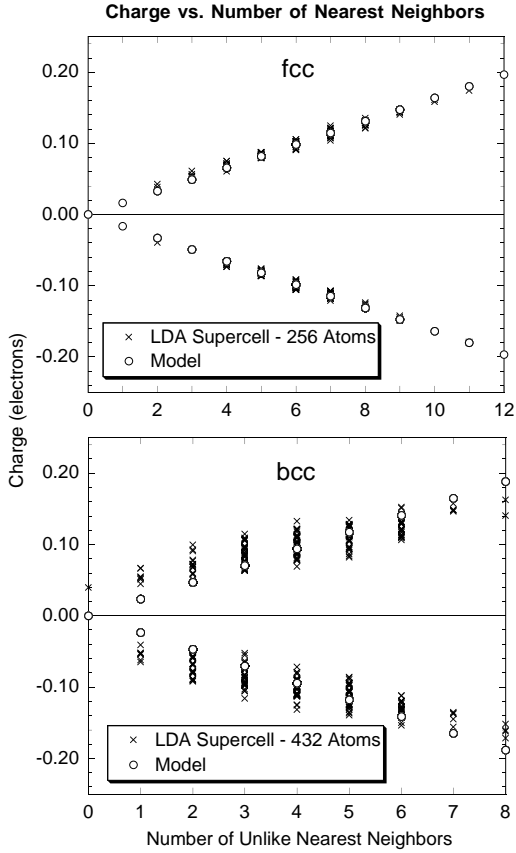}\hfil}
\nobreak\bigskip
\caption{Charge versus number of unlike nearest neighbors.
Shown are the predictions of the charge model of 
Eq. (\protect\ref{charge}) and the
results of LDA supercell calculations of 
Ref. \protect\onlinecite{Faulkner.conf}.
Values of $\lambda$ [Eq. (\protect\ref{lambda.lda})]
were obtained by a least-squares fit to the 
LDA supercell data.}
\label{fig.charge}
\end{figure}

The prediction of the model of Eq. (\ref{charge})
for the dependence of charge
on the number of unlike-nearest neighbors is clear:  It
is a linear relation.
The charges predicted by this simple model (open circles)
are compared with those obtained in extensive 
LDA supercell calculations (crosses) \cite{Faulkner.conf} in 
Fig. \ref{fig.charge}.
A least-squares fit to the LDA supercell data
gives values of the parameter $\lambda$:
\begin{equation}
\label{lambda.lda}
\lambda^{\rm fcc} = 0.00819, \; \; \;
\lambda^{\rm bcc} = 0.01176.
\end{equation}
The LDA calculations demonstrate that 
(i) the linear prediction
of the model is accurate for fcc alloys, but
(ii) for bcc alloys, as recognized by 
Faulkner {\em et al.}, \cite{Faulkner95,Faulkner.conf} 
there are fluctuations about the straight line.
Recall that in an fcc structure the distances from the origin
to the $n^{\rm th}$ shell are $R_1$, $1.414R_1$, and $1.732R_1$
for $n$=1,2,3, while for bcc these distances are
$R_1$, $1.155R_1$, and $1.633R_1$. Thus, the bcc structure
shows a weaker distinction
between first and second neighbors.
We consider below (Section \ref{generalizations})
possible generalizations of the charge model 
for bcc alloys which account for these fluctuations
by extending the linear relation (\ref{charge}) to
more than one coordination shell.

\subsection{Relation between Constant-Occupation-Averaged Charge 
and Coulomb Shift}

\jot=20pt
\narrowtext
\begin{table*}[htb]
\caption{
Comparison of physical properties of random alloys
which are a consequence of 
(a) the charge model of Eq. (\protect\ref{charge}), 
(b) the generalized charge model of Eq. (\protect\ref{charge.gen}), and 
(c) those obtained from LDA supercell simulations of
Refs. \protect\onlinecite{Faulkner95,Faulkner.conf}.
In cases (a) and (b), we assign charges to sites according to a
given model [Eq. (\protect\ref{charge}) or (\protect\ref{charge.gen})] and
then calculate the Coulomb shift $V_i$ [Eq. (\protect\ref{shift})]
by applying the Ewald method to the assigned charges.}
\label{compare}
\begin{tabular}{cccc}
&$R_{\rm eff}/R_1$&$\gamma/R_1$ [Ry$^{-1}*$a.u.$^{-1}$]&
$\langle E_M \rangle_R$ [mRy/atom]\\
\tableline
\\
\multicolumn{4}{c}{fcc ($x$=1/2)}\\
\\
Model - Analytic          &1.00&-0.132&-2.60\\
Model - 256 atoms         &1.06&-0.130&-2.52\\
Model - 16 atom SQS       &0.99&-0.134&-2.60\\
LDA supercell - 256 atoms &0.97&-0.123,-0.127&-2.61\\
Gener. model - 256 atoms  &1.02&-0.120&-2.55\\
Gener. model - 16 atom SQS&1.03&-0.118&-2.52\\
\tableline
\\
\multicolumn{4}{c}{bcc ($x$=1/2)}\\
\\
Model - Analytic         &1.00&-0.163&-2.57\\
Model - 432 atoms        &0.98&-0.155&-2.64\\
LDA supercell - 432 atoms&1.02&-0.114,-0.116&-2.67\\
Gener. Model - 432 atoms &1.19&-0.119&-2.34\\
\end{tabular}
\end{table*}

Both the charge model of Eq. (\ref{charge}) and the LDA supercell
calculations result in a relationship 
between the constant-occupation-averaged charges $\langle q \rangle_A$
and $\langle q \rangle_B$ [Eq. (\ref{charge.rand})] and the
constant-occupation-averaged Coulomb shifts 
$\langle V \rangle_A$
and $\langle V \rangle_B$ [Eq. (\ref{shifta})]
of the form:
\begin{equation}
\langle V \rangle_{A,B} = \frac{-\langle q \rangle_{A,B}}{R_{\rm eff}}
\end{equation}
According to Eq. (\ref{qave.Vave}), the charge model of Eq. (\ref{charge})
gives $R_{\rm eff} = R_1$
(where $R_1$ is the nearest neighbor distance).
Values of $R_{\rm eff}/R_1$ 
for the charge model and for the LDA supercell calculations
are compared in Table \ref{compare}.
The simple charge model of Eq. (\ref{charge}) reproduces
$R_{\rm eff}/R_1$
of the LDA supercell calculations 
(0.97 and 1.02 for fcc and bcc, respectively) to within a few percent.

In Eq. (\ref{charge.rand}), it is shown that the simple charge
model predicts that the difference $\Delta$ in 
constant-occupation-averaged charges
is independent of composition.  The LDA supercell 
results (Fig. 3 of Ref. \onlinecite{Faulkner95}) 
also show that $\Delta$ depends very little on composition,
in agreement with the model prediction.

The difference between constant-occupation-averaged 
charges $\Delta(\eta)$
for the charge model was also 
derived as a function of LRO parameter in Eq. (\ref{charge.lro}).
The charge model prediction is that $\Delta(\eta)$ should increase
for the LRO $B2$ alloy relative to the random alloy by a 
factor $\Delta^{B2}(1)/\Delta(0) = 2$.  
The LDA calculations
\cite{Faulkner95} show that the introduction of
LRO does increase $\Delta$ 
from a value of 0.20066 for
the random alloy to 0.25178 for the $B2$ ordered alloy,
giving $\Delta^{B2}(1)/\Delta(0) = 1.25$, somewhat smaller
than but qualitatively consistent with 
the prediction of the charge model.
In considering the disparity between the magnitude of
$\Delta^{B2}(1)/\Delta(0)$ of the simple charge
model and that of LDA, one should remember that in
the latter, point charges are defined by a non-unique
partitioning of space.

The influence of SRO on $\Delta$ was derived for the
charge model in Eq. (\ref{charge.sro}) where it was
shown that ordering type SRO (as found in Cu-Zn alloys)
should increase $\Delta$ relative to the random alloy.
The introduction of SRO in the LDA supercell 
calculations \cite{Faulkner95}
increases $\Delta$ from 0.20066 for a random simulation
to 0.20554 in a simulation with some degree of SRO.  This
increase is again consistent with the
predictions of the charge model.

\subsection{The $q-V$ Relation between Charges and Coulomb Shifts}

The large supercell LDA calculations find a 
linear relationship between the charges $q_i$ on individual
sites and the Coulomb shifts $V_i$ on those sites.
The simple charge model of Eq. (\ref{charge})
predicts [Eq. \ref{qV.analytic}] a linear behavior between
charge and constant-charge-averaged Coulomb
shift in agreement with the LDA supercell calculations.
The slope $\gamma$ of this linear relation is compared
with the slopes from the LDA supercell calculations in Table \ref{compare}.
(Note that both charge and Coulomb shift are proportional to
the parameter $\lambda$ of the model, and thus,
the slope $\gamma$ is independent of $\lambda$.)
The relative slope $\gamma/R_1$ of the model (-0.132) is within a
few percent of the LDA supercell results for fcc alloys 
(-0.125 $\pm$ 0.002), while
for bcc alloys, the slope of the model 
(-0.163) is too large in magnitude relative to the LDA
result (-0.115 $\pm$ 0.001).

%
%
\begin{figure*}[htb]
\hbox to \hsize{\epsfxsize=0.80\hsize\hfil\epsfbox{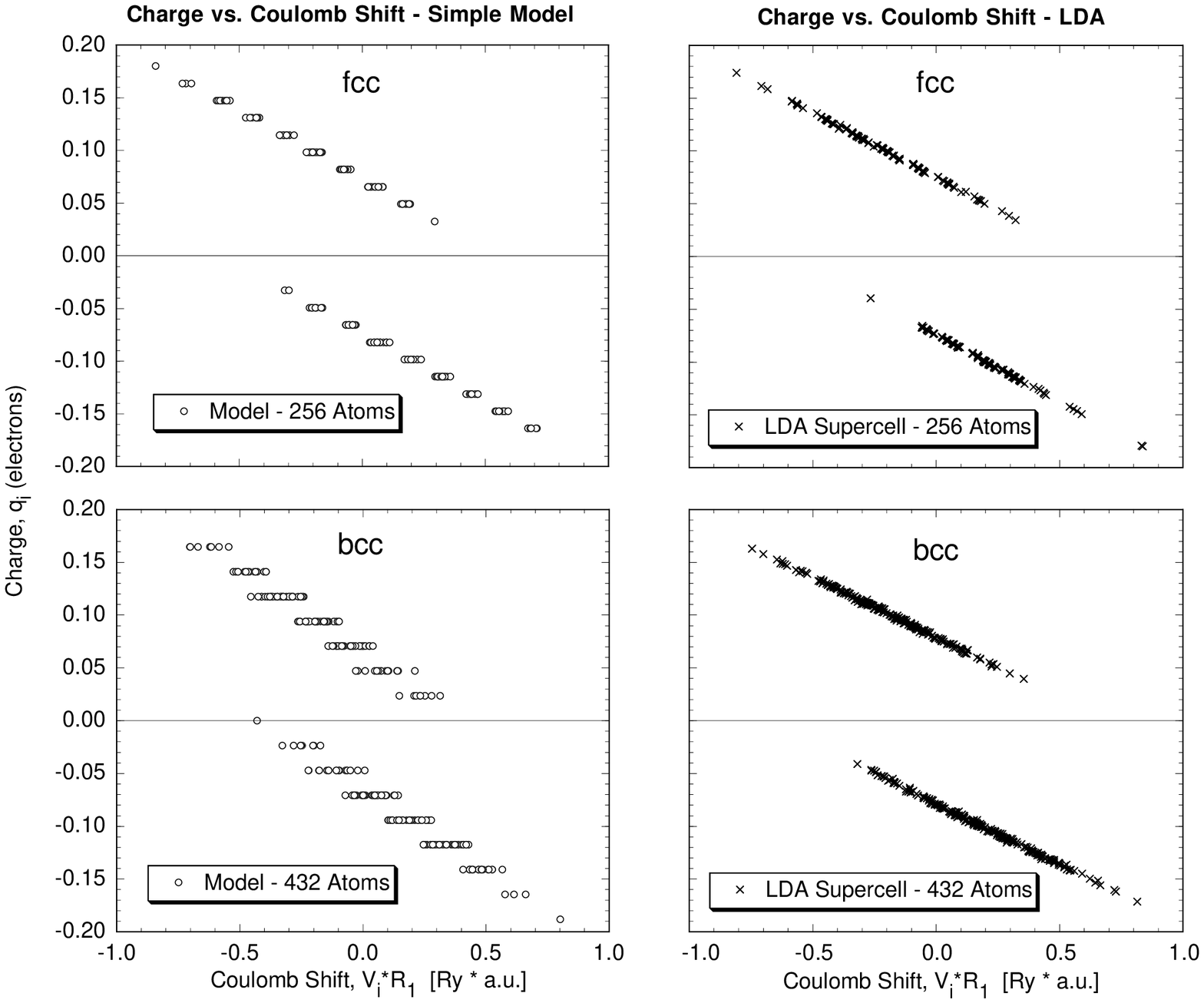}\hfil}
\nobreak\bigskip
\caption{Charges versus Coulomb Shifts as predicted by the
charge model of 
Eq. (\protect\ref{charge}) using the values 
of $\lambda$ and $R_1$
given in Eqs. (\protect\ref{lambda.lda})
and (\protect\ref{r1.lda}).
Also shown are the
results of LDA supercell calculations of 
Ref. \protect\onlinecite{Faulkner.conf}.}
\label{qV}
\end{figure*}

The relationship between $q_i$ and the distinct Coulomb shift $V_i$
(not the constant-charge-averaged $\overline{V}_i$) as obtained
in the simple model of Eq. (\ref{charge}) is
shown in Fig. \ref{qV},
where it is
contrasted with the results of the LDA calculations 
of Ref. \onlinecite{Faulkner.conf}.
The fluctuations in Coulomb shift
about the average linear behavior of $q_i$ and
$V_i$ are quite small in the fcc random alloy,
but are substantial in the bcc alloy.
We have next determined the effect of these fluctuations on the
electrostatic energy $\langle E_M \rangle_R$ 
of the random alloy:  If the linear relation Eq. (\ref{qV.analytic})
between charge and Coulomb shift (neglecting fluctuations)
is used in
Eq. (\ref{e.point.2}) to compute the random alloy energy, we recover 
precisely the same energy {\em including fluctuations} derived in
Eq. (\ref{e.rand}).  Thus, although the fluctuations in $V_i$ are
graphically impressive (Fig. \ref{qV}), 
{\em the energetic consequence of these fluctuations
is strictly zero}, simply indicating that the fluctuations in
Coulomb shift are symmetrical about the average linear behavior.

\subsection{Charge-Charge Correlations Functions}

%
%
\begin{figure*}[tb]
\hbox to \hsize{\epsfxsize=0.80\hsize\hfil\epsfbox{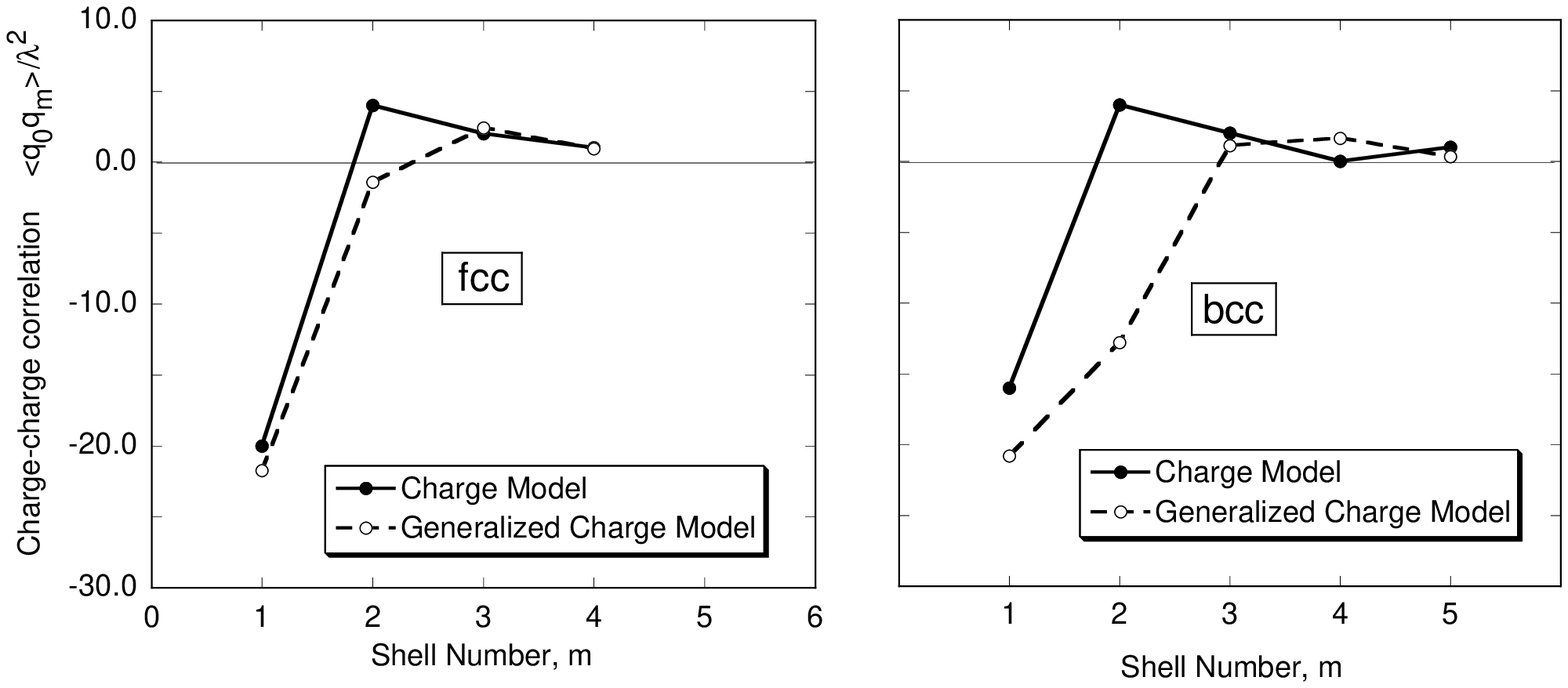}\hfil}
\nobreak\bigskip
\caption{Charge-charge correlation functions for
random $x$=1/2 alloys.  Shown are the correlations
of the simple charge model and the generalized charge model for 
fcc and bcc lattices.  
For the simple charge model, the analytic results of
Eq. (\protect\ref{corr.random}) are plotted while the
correlations of the fcc (bcc) generalized charge model are obtained
from a 16,384 (8,192)-atom simulation, configurationally averaged 
over 20 configurations.
Correlation functions are shown
normalized to $\lambda_1^2$}
\label{charge.charge}
\end{figure*}

The simple charge model
predicts specific values for the charge-charge correlations
of random alloys
given in Eq. (\ref{corr.random}).
The quantitative results of the LDA supercell for these correlations are
a bit unclear:  In Ref. \cite{Faulkner.conf}, the authors
note that for 256-atom LDA supercell calculations, the
nearest-neighbor correlations are sizeable, but they also note
that the values beyond
the nearest-neighbor shell are smaller than the predictions of
the model, although these values are not too well known due to the
relatively small size of the simulation cell.  When larger
LDA supercell simulations become available, a comparison of charge-charge
correlations from LDA supercell with the predictions of 
Eq. (\ref{corr.random}) (and those of the generalized charge model 
described below) would be of interest.
The analytic values of the charge-charge correlation functions
of Eq. (\ref{corr.random})
are plotted in Fig. \ref{charge.charge}.  We have also compared
these analytic values with those obtained from our large-unit-cell
simulations of the charge model (not shown).  
Although the correlations for
the nearest neighbor shell are robust with respect to unit cell size,
the correlations
for the more distant 3rd, 4th, and 5th shells are extremely
sensitive to the size of the simulation cell:  For a single fcc 256-atom
simulation, one can even find 3rd and 4th neighbor correlations which
have opposite sign relative to the exact analytic values.  Even for
very large (16384-atom) fcc simulations configurationally averaged over
20 configurations, the 3rd and 4th neighbor correlations may differ from
the analytic values by $\sim10\%$.  Thus, in order to compare
the LDA charge-charge correlations for random alloys with the 
analytic results of the simple charge model,
the size of the LDA supercell calculations would have to be 
significantly increased.

\subsection{Coulomb Energy of Random Alloys}

The Madelung energy of the simple charge model for a random alloy 
is given in Eq. (\ref{e.rand}) in terms of the parameter $\lambda$ 
and the nearest-neighbor distance $R_1$.  If we use the numerical
values for $\lambda$ given in Eq. (\ref{lambda.lda}) and the
nearest neighbor distances used in the LDA supercell calculations
for Cu-Zn alloys, 
\begin{equation}
\label{r1.lda}
R_1^{\rm fcc}=4.879{\rm a.u.}; \; \; R_1^{\rm bcc}=4.763{\rm a.u.},
\end{equation}
we obtain  the electrostatic energies of the simple charge model
for Cu-Zn:
\jot=0pt
\begin{eqnarray}
\langle E_M \rangle_R({\rm fcc};x=0.5) &=& -2.60 \; {\rm mRy/atom}, 
\nonumber \\
\langle E_M \rangle_R({\rm fcc};x=0.7) &=& -2.18 \; {\rm mRy/atom}, 
\nonumber \\
\langle E_M \rangle_R({\rm bcc};x=0.5) &=& -2.57 \; {\rm mRy/atom}.  
\end{eqnarray}
These values are compared with the Cu-Zn LDA supercell values 
\cite{Faulkner.conf} in Table \ref{compare}:
\begin{eqnarray}
\langle E_M \rangle_R({\rm fcc};x=0.5) &=& -2.61 \; {\rm mRy/atom}, 
\nonumber \\
\langle E_M \rangle_R({\rm fcc};x=0.7) &=& -2.20 \; {\rm mRy/atom}, 
\nonumber \\
\langle E_M \rangle_R({\rm bcc};x=0.5) &=& -2.67 \; {\rm mRy/atom}.  
\end{eqnarray}
For all cases, the prediction of the simple model is extremely
accurate:  the model energies fall within 0.1 mRy of the 
LDA supercell calculations.  
Although the model of Eq. (\ref{charge}) was shown to have significant
fluctuations in the $q-V$ relations (Fig. \ref{qV}), 
{\em these fluctuations
have a vanishing contribution to the Coulomb energy}, and thus the model
produces accurate energetics.

\subsection{Approximating Large Random Supercells by Small-Cell
``Special Quasi-Random'' Structures}

Our foregoing discussions were based on (either LDA or Ewald)
simulations of rather large supercells (e.g., 256-432 atom).
We next examine the extent to which {\em specially selected}
small cells can mimic larger, non-specially selected cells.
Special quasi-random structures (SQS) \cite{SQS} are small unit-cell
structures which are constructed in such a way so that structural
(not charge-charge) correlation functions $\overline{\Pi}_{\rm SQS}$
match as closely as possible those of the random alloy 
($\overline{\Pi}_{\rm SQS} \sim 
\overline{\Pi}_{\rm Random}$) for several neighboring
shells.  In this way the SQS is a small-unit-cell ordered structure
which mimics the random alloy.
It is interesting to see how the charge model calculation of a 
small-unit-cell SQS compares with the large scale 256-atom
simulations described above.

We have performed an Ewald calculation for a 16-atom, fcc-based SQS 
structure (denoted SQS-16) with
point charges taken from the charge model of Eq. (\ref{charge}).  
Structural information for SQS-16 is given in Appendix B.
The resulting $R_{\rm eff}/R_1$,
$\gamma/R_1$, $\langle E_M \rangle_R$,
and the $q-V$ relation for the SQS-16 are collected
in Table \ref{compare}, where they are compared with
analogous calculations using a (randomly selected) 256-atom cell.
We see that the 16-atom SQS calculation matches
the 256-atom simulation for all properties to within
a few percent.  
Also, the electrostatic energy of the SQS-16
$\langle E_M \rangle_{\rm SQS}/E_0 = -0.740$ 
compares much more favorably with the exact value of
-0.7395 than does the energy of the 256-atom simulation
$\langle E_M \rangle_R^{\rm 256-atom}/E_0 = -0.716$.  Thus,
for the case of electrostatic energies, the 16-atom SQS
provides a {\em more accurate} depiction of the random alloy
than does a single, randomly-selected 256-atom configuration.
It would be interesting to compare the
LDA energies of the SQS-16 with those of larger, but randomly selected
supercells. \cite{Faulkner95}

\section{Generalizations of the Model}
\label{generalizations}

\subsection{Summary of Successes and Failures of the Simple Model}

We have thus far ascertained the physical predictions of
the simple, nearest-neighbor charge model of Eq. (\ref{charge}), 
and compared them with the results of large
LDA supercell calculations of Ref. \onlinecite{Faulkner95,Faulkner.conf}.
In many cases, the simple charge model accurately predicts the
electrostatic properties of LDA:  

(i) The behavior of
charge versus nearest neighbor environment is reproduced well
by LDA calculations of fcc-based alloys (Fig. \ref{fig.charge}a).

(ii) The proportionality $R_{\rm eff}/R_1$ 
between constant-occupation-averaged charge 
$\langle q \rangle_{A,B}$ and Coulomb
shift $\langle V \rangle_{A,B}$
of the model is the same as that of LDA to within a
few percent (Table \ref{compare}).

(iii) For fcc alloys, the linear $q-V$ relation of LDA is
well reproduced (including fluctuations) by the model (Fig. \ref{qV}).  
The model value for the slope $\gamma/R_1$
of the $q-V$ relation in fcc alloys
is within a few percent of LDA (Table \ref{compare}).

(iv) The Coulomb energies of the model are extremely accurate
with respect to the LDA values (to within 0.1 mRy/atom).

(v)  The slope of the $q-V$ relation, $\gamma$,
in Fig. \ref{qV} are the same for $A$ and $B$.  
The LDA supercell calculations
also show similar slopes ($\gamma/R_1$) for charges on
Cu ($-0.123$) or Zn ($-0.127$) atoms in the fcc $x$=1/2 alloy, 
or for 
Cu ($-0.114$) or Zn ($-0.116$) atoms in the bcc $x$=1/2 alloy.

(vi)  The slope of the charge {\em versus} number of unlike 
nearest neighbors (Fig. \ref{fig.charge}) are
negatives of one another.  LDA supercell calculations (for fcc
alloys) support this (Fig. \ref{fig.charge}).  

(vii)  In the impurity
limit, the model predicts 
that the charge on $A$ embedded in
pure $B$ is equal (in magnitude) to that of $B$ embedded in pure
$A$,
\begin{equation}
\label{dilute}
| \langle q \rangle_A(x\rightarrow1) | =
| \langle q \rangle_B(x\rightarrow0) | = 2Z\lambda.  
\end{equation}
The LDA supercell calculations also show this behavior
(see Fig. \ref{fig.charge}), 
for an atom surrounded completely by unlike neighbors.
Note that neither the simple model 
nor the LDA
supercell simulations include the effects of
atomic relaxations, which could likely
eliminate the degeneracy 
of Eq. (\ref{dilute}).
[To describe relaxed configurations, it is anticipated that
more parameters (e.g., bond lengths) would need
to be introduced into the model.]

(viii) $\lambda$ is composition-independent in the
charge model; values of $\lambda$ 
(Table I of Ref. \cite{Faulkner.conf})
extracted from the LDA supercell calculations 
also demonstrate that $\lambda$ 
is not sensitive to concentration.
We reiterate that the charge model describes only
unrelaxed configurations at a fixed volume. 
For lattice-mismatched systems, alloys of different composition
will have different volumes, and the charge transfer will
depend on this volume.  To model this effect, $\lambda$
should be explicitly volume-dependent.  This {\em explicit}
volume-dependence would lead to an {\em implicit} dependence
of $\lambda$ on composition.  (Presumably, this 
implicit composition-dependence is not seen in the
LDA supercell data of Ref. \cite{Faulkner.conf} due to 
the fact that the system studied, Cu-Zn, has a relatively
small lattice-mismatch.)  However, this should not
be confused with an {\em explicit} composition-dependence
of $\lambda$.

Although there are many cases of agreement between the predictions of
the charge model [Eq. (\ref{charge})] and the electrostatics of large
LDA calculations, certain discrepancies arise in these comparisons:

(i) The LDA calculations
show that the charge is not a single-valued function when
plotted versus the number of unlike nearest neighbors
(Fig. \ref{fig.charge}).  Although there is not much
width to the distribution for fcc alloys, there is a
significant width for bcc alloys.
Also demonstrated by Fig. \ref{fig.charge} is that charges
in the model of Eq. (\ref{charge}) are quantized since
the number of unlike nearest neighbors must be an integer.
The LDA calculations (particularly for bcc) show no
such quantization.

(ii) The slope of the $q-V$ relation ($\gamma/R_1$) 
for bcc alloys (Table \ref{compare}) is 
significantly larger in magnitude in the model (-0.163) than
in the LDA calculations (-0.115).

(iii) There are significant fluctuations about a
linear $q-V$ relation
obtained by the charge model; however, the LDA calculation
show a nearly perfect linear relation with no fluctuations
(Fig. \ref{qV}).
The fluctuations of the charge model are especially pronounced
for bcc alloys.

\subsection{Generalizing the Model}

The charge model of Eq. (\ref{charge}) is based on 
the obvious chemical fact that atomic charge results
from charge {\em transfer}, and that the latter depends
on the identity of the {\em neighbors}, since charge transfer
does not occur between chemically equivalent sites.
Thus, $q_i$ should depend on the local environment of site $i$.
Magri {\em et al.} \cite{Magri90} took 
first neighbors to be the ``leading order'' contribution
to the local environment, and for the case that Magri
{\em et al.} treated - fcc alloys - we have seen that
the charge model provides an adequate description of
electrostatics.  However, alloys based on different
lattice types can have different structural environments,
in terms of coordination numbers and neighbor distances:
In the fcc lattice there is a significant
``gap'' between the distance of the first coordination
and that of the second.
In bcc, however,  the ``gap'' is after the second shell.
This suggests that one
generalization of the charge model which would 
affect bcc and fcc alloys differently
is to allow the charges in the model
to be dependent on more distant neighbor shells.
Thus, instead of requiring the charges to be a function of the
number of unlike {\em nearest} neighbors, we define a
{\em generalized charge model} in which charges are a function of
the number of unlike nearest neighbors 
{\em on the first several shells} $s$ of neighbors:
\begin{equation}
\label{charge.gen}
q_i = \sum_s \lambda_s \sum_{k_s=1}^{Z_s} [ \hat{S}_i - \hat{S}_{i+k_s} ],
\end{equation}
For this generalized charge model, the charge on a site $i$
is linearly proportional to the generalized number of 
neighbors, $\tilde{N}$:
\begin{equation}
\label{nn.gen}
\tilde{N} = \sum_s N_i^{(s)} \frac{\lambda_s}{\lambda_1}
\end{equation}
where $N_i^{(s)}$ is 
the number of unlike neighbors in the $s$th shell
for the atom at site $i$. 
In this new ``generalized'' charge model of Eq. (\ref{charge.gen}),
there are $S$ parameters, where $S$ is the number of
shells included.

To determine the parameters of the generalized charge
model, we have fit (via a least-squares procedure) 
the charges of the LDA supercell
calculations to Eq. (\ref{charge.gen}) including
five shells.  The parameters
$\lambda_s$ are zero, for all intents and purposes,
for $s > 2$ in fcc and $s > 3$ in bcc.  
Within these ranges, we found 
\begin{eqnarray}
\label{lambda.lda.gen}
&&\lambda_1^{\rm fcc} = 0.00745, \; \;
\lambda_2^{\rm fcc}/\lambda_1^{\rm fcc} = 0.214, \; \;
\lambda_{n>2}^{\rm fcc}/\lambda_1^{\rm fcc} \sim 0
\nonumber \\
\\
&&\lambda_1^{\rm bcc} = 0.00786, \; \;
\lambda_2^{\rm bcc}/\lambda_1^{\rm bcc} = 0.660, \; \; \nonumber \\
&&\lambda_3^{\rm bcc}/\lambda_1^{\rm bcc} = 0.0645, \; \;
\lambda_{n>3}^{\rm bcc}/\lambda_1^{\rm bcc} \sim 0. \; \;
\end{eqnarray}

\subsection{Testing the Generalized Model}

%
%
\begin{figure}[tb]
\hbox to \hsize{\epsfxsize=0.80\hsize\hfil\epsfbox{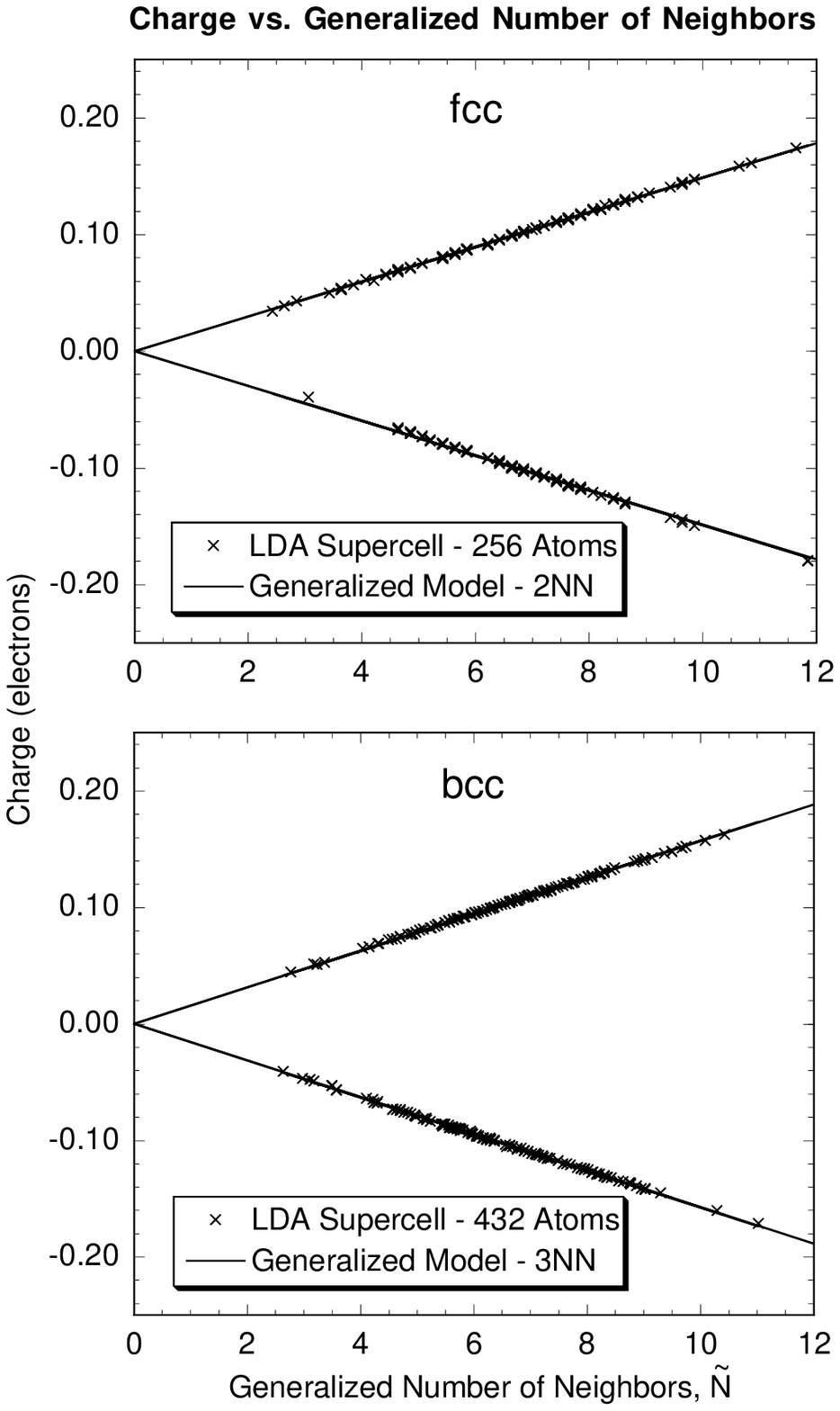}\hfil}
\nobreak\bigskip
\caption{Charge versus number of generalized 
neighbors $\tilde{N}$ Eq. (\protect\ref{nn.gen}).
Shown are the predictions of the generalized 
charge model of 
Eq. (\protect\ref{charge.gen}) using the values 
of $\lambda_s$ fit to large-unit-cell LDA calculations
given in Eq. (\protect\ref{lambda.lda.gen}).
Also shown are the
charges of the LDA large-unit-cell calculations of 
Ref. \protect\onlinecite{Faulkner.conf}.}
\label{fig.charge.gen}
\end{figure}

%
%
\begin{figure*}[htb]
\hbox to \hsize{\epsfxsize=0.80\hsize\hfil\epsfbox{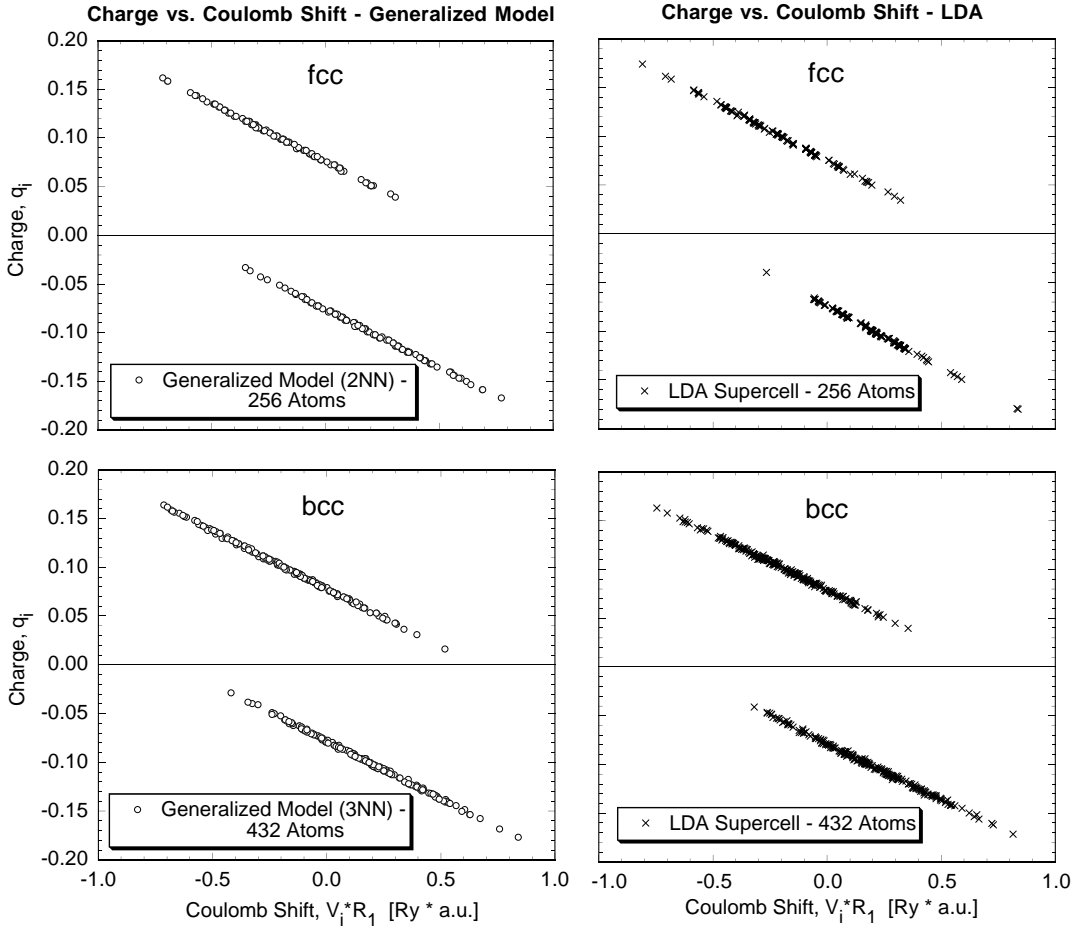}\hfil}
\nobreak\bigskip
\caption{Charge versus Coulomb shift as predicted
by the generalized 
charge model of 
Eq. (\protect\ref{charge.gen}) using the values 
of $\lambda_s$ and $R_1$ given in 
Eqs. (\protect\ref{lambda.lda.gen})
and (\protect\ref{r1.lda}).}
\label{qV.gen}
\end{figure*}

We show in Figs. \ref{charge.charge}, \ref{fig.charge.gen},
and \ref{qV.gen}
results for generalized fcc (bcc)
charge models
including the first two (three) neighbors shells with
these values of $\lambda_s$.

Figure \ref{charge.charge} shows the charge-charge correlations
of random $x$=1/2 alloys
predicted by the simple charge model of Eq. (\ref{charge}),
and the generalized charge model of Eq. (\ref{charge.gen}).
Although there are currently no LDA results with which
to compare (due to the size of the current LDA supercells), 
we note that the generalized charge model
changes the sign of the second neighbor correlation in
both fcc and bcc relative to the simple model.  
It would be interesting to compare these
correlations with those of LDA when larger supercell calculations
become available.

The charge $q_i$ {\em versus} the generalized 
number of neighbors $\tilde{N}$ is shown in Fig. \ref{fig.charge.gen}
for LDA and for the model of Eq. (\ref{charge.gen}).
For fcc alloys, the corrections induced by generalizing
the charge model are small since the original model of
Eq. (\ref{charge}) is already very good.  
The predictions of the generalized charge
model fit the LDA supercell data 
extremely well even for bcc alloys, where the nearest-neighbor
model of Eq. (\ref{charge}) was lacking.  

\begin{figure}[tb]
\hbox to \hsize{\epsfxsize=0.80\hsize\hfil\epsfbox{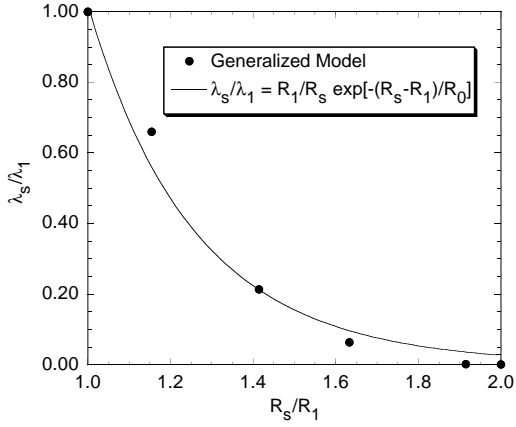}\hfil}
\nobreak\bigskip
\caption{Charge transfer parameters $\lambda_s$ 
of generalized model
of Eq. (\protect\ref{charge.gen}) as a function of
distance.  Also shown is a fit to the parameters to 
the exponential function of Eq. (\protect\ref{screen}).  
The fitted value is $R_0 = 0.34R_1$.}
\label{lambda.vs.r}
\end{figure}

Figure \ref{qV.gen} shows
the relation between charge $q_i$ and Coulomb shift $V_i$
of the generalized charge model, comparing the results
with LDA. 
Like LDA, the generalized charge
model predicts a linear relation between $q_i$ and $V_i$
with almost no fluctuations.  Furthermore, the slope of 
these linear relations are in excellent agreement with
the LDA supercell data (Table \ref{compare}),
provided that cutoffs for fcc and bcc are at second and third neighbor 
shells, respectively.  
Thus, the generalized charge model of Eq. (\ref{charge.gen})
rectifies all of the discrepancies noted above 
(Section \ref{generalizations}A)
between model and LDA calculations.
[The fcc model for 
nearest-neighbors only is already accurate with
respect to LDA calculations (Figs. \ref{fig.charge} and
\ref{qV}), thus generalizing the fcc charge model
to first- and second-neighbors does not produce a 
large effect.]
In Fig. \ref{lambda.vs.r}, we show the values of the
parameters $\lambda_s$ versus distance of the shell $s$.
One can see that the parameters are reasonably well fit by
an exponential function
\begin{equation}
\label{screen}
\lambda_s = \frac{\lambda_1 R_1}{R_s}e^{-(R_s-R_1)/R_0}
\end{equation}
with a decay length of $R_0 = 0.34R_1$.
This suggests that in an alloy the net charge on each
site is screened effectively in a very short range.

%
%

Since the generalized charge model predicts a linear
$q-V$ relation in disordered alloys, with almost no fluctuations, 
one can also obtain a generalized model of the {\em Coulomb
shifts} in an alloy
\begin{equation}
\label{shift.gen}
V_i \propto \gamma/R_1 
(\sum_s \lambda_s \sum_{k_s=1}^{Z_s} [ \hat{S}_i - \hat{S}_{i+k_s} ]).
\end{equation}
Thus, the Coulomb shifts, like
the charges, depend only on the occupation of the
first few neighboring shells.

\subsection{Extracting Values of $\lambda$ from LDA:  
Supercell-Size-Dependence}

%
%
\begin{figure}[tb]
\hbox to \hsize{\epsfxsize=0.80\hsize\hfil\epsfbox{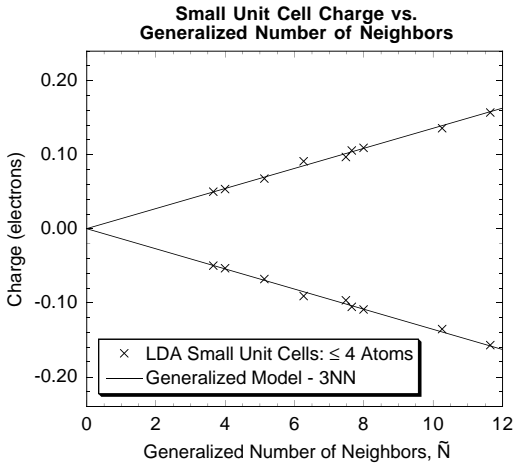}\hfil}
\nobreak\bigskip
\caption{Charge versus number of generalized
neighbors $\tilde{N}$ Eq. (\protect\ref{nn.gen}).
Shown are the predictions of the generalized
charge model of
Eq. (\protect\ref{charge.gen}) using the values
of $\lambda_s$ fit to small-unit-cell data
given in Eq. (\protect\ref{lambda.lapw.gen}).
Also shown are the
charges of the LAPW small-unit-cell calculations of
the present work.}
\label{fig.charge.lapw}
\end{figure}

We have demonstrated the validity of the generalized model
of point charges and shown how the parameters of the model $\lambda_s$
may be extracted from large-unit-cell LDA calculations.  However, the
models of point charges (both the simple and generalized models) assume
that the {\em physical mechanism} underlying excess charge on a site is the
same for ordered and random alloys.  
This suggests that the values of $\lambda_s$ could be obtained from
small-unit-cell calculations.
For computational simplicity, one should know whether it is
equally valid to extract
values of $\lambda_s$ from ordered or random alloys, and whether
one can use LDA calculations of small cells ($\sim$2-4 atoms) to
extract the values of $\lambda_s$.
To this end, we have complemented the large unit-cell LDA calculations
of Faulkner {\em et al.} \cite{Faulkner95} on random bcc Cu-Zn alloys
by performing calculations of several {\em ordered small-unit-cell} 
bcc-based Cu-Zn ordered compounds.
We use the linearized augmented plane wave (LAPW) method.  \cite{singh} 
The ordered structures considered are all bcc superlattices: 
Cu$_1$Zn$_1$ (001), Cu$_2$Zn$_2$ (111), Cu$_2$Zn$_2$ (001), 
Cu$_2$Zn$_1$ (001), Cu$_3$Zn$_1$ (111), and Cu$_1$Zn$_1$ (101).
All of these compounds have 2-4 atoms/cell and the first five are
commonly referred to by their Structurbereicht designations:  $B2$, $B32$,
$B11$, $C11_b$, and $D0_3$, respectively.
In the LAPW calculations, we have used the
exchange correlation of Wigner. \cite{wigner}
The muffin-tin radii are chosen to be equal (2.2 a.u.) for both Cu and Zn.
Brillouin-zone integrations are performed using the
equivalent {\bf k}-point sampling method \cite{froyen}.
Since the charge model is appropriate only for charges in
unrelaxed geometries at
fixed volume, all computations were done in ideal geometries
at a single volume ($a$=5.56 a.u.), 
even though several compositions are considered.  
The excess charges were extracted from
the LAPW calculations by integrating the charge density inside
the muffin-tin spheres and dividing the interstitial charge evenly
between the atoms in the unit cell.
%

The LAPW charges for the six small-unit-cell compounds 
calculated were fit to a form
of the generalized charge model of Eq. (\ref{charge.gen}) with 1st-3rd
neighbor shells.  The parameters of the generalized model fit
to these small unit-cell calculations,
\begin{equation}
\label{lambda.lapw.gen}
\lambda_1^{\rm bcc} = 0.00680, \; \;
\lambda_2^{\rm bcc}/\lambda_1^{\rm bcc} = 0.609, \; \;
\lambda_3^{\rm bcc}/\lambda_1^{\rm bcc} = 0.131, \; \;
\end{equation}
agree well with those fit to 
large unit-cell data [Eq. (\ref{lambda.lda.gen})]: \cite{note.lapw}  
The parameters of Eq. (\ref{lambda.lapw.gen}) fit to small-unit-cell
LDA calculations lead to a $q-V$
relation which is linear, with no fluctuations, and has a 
slope of $\gamma/R_1$=0.112, compared with $\gamma/R_1$=0.119
for the parameters of the generalized model fit to 
large-unit-cell LDA data. 
The charges extracted from small-unit-cell LDA calculations 
are shown in Fig. \ref{fig.charge.lapw} as a 
function of generalized number of neighbors $\tilde{N}$
[using the values of $\lambda_s$ in Eq. (\ref{lambda.lapw.gen})].
These calculations demonstrate that the parameters of the generalized
model may be determined from calculations of 
several small-unit-cell ordered compounds in unrelaxed geometries
at fixed volume.  
If one wishes to assess the explicit volume-dependence of the
parameters, one only needs to repeat these types of calculations
at a few different volumes.  We have performed such volume-dependent
calculations at $a$=5.36 and $a$=5.75 (in addition to the $a$=5.56
calculations described above), and find that the values of 
$\lambda$ only have a slight volume-dependence in this range:  The
value of $\lambda_1$ at $a$=5.36 is about 6$\%$ larger in magnitude
than $\lambda_1$ at $a$=5.75.  Also, in this volume range the ratios 
$\lambda_2/\lambda_1$ and $\lambda_3/\lambda_1$ vary by less than 
their uncertainty due to the fit.

\section{Conclusions}

Recent \cite{Faulkner95,Faulkner.conf}
large scale (256-432 atom) local
density approximation (LDA) supercell
calculations of Cu$_{1-x}$Zn$_x$ random alloys
allow us to examine the adequacy of simple models
describing the dependence of point charges in disordered
alloys on the atomic environment.
We find that a model 
in which the excess charge $q_i$ on an atom in an ordered
or random alloy depends linearly on the number $N_i^{(1)}$ of
unlike neighbors in its first coordination shell
correctly describes
the trends in charge versus number of unlike nearest neighbors,
(particularly for fcc alloys),
the magnitudes of Coulomb energies
in random Cu$_{1-x}$Zn$_x$ alloys, and
the relationships between {\em constant-occupation-averaged} 
charges $\langle q_i \rangle$ and Coulomb
shifts $\langle V_i \rangle$ in the random alloy.
However, for bcc alloys the {\em fluctuations} predicted 
by the model in the $q_i$ {\em vs} $V_i$
relation exceed those found in the LDA supercell
calculations.
Although we found that the fluctuations present in
the model have a vanishing contribution
to the electrostatic energy,  generalizing
the bcc (fcc) model to include a dependence of the charge on
the atoms in the first {\em three (two) shells}
(rather than the first shell only)
removes the fluctuations from the model,
in complete agreement with the LDA data.

Other possible generalizations of the charge model include:
(i) non-linearities in the charge as a function of number of 
neighbors and
(ii) charges which depend not only on the {\em number} of
nearest neighbors, but also on the particular arrangement
of the neighbors.  This type of dependence would lead to
not only pair correlations among charges, but also 
multibody correlations.  
Currently, there are no indications that these types
of generalizations are warranted.

We thank Drs. Y. Wang and G. M. Stocks for providing us their
LDA data of Ref. \cite{Faulkner95}. 
We also thank Dr. Z.-W. Lu for insightful discussions.
This work was supported by the Office of Energy Research
(OER) [Division of Materials Science of the Office of Basic Energy
Sciences (BES)], U. S.  Department of Energy, under contract No.
DE-AC36-83CH10093.

\begin{center}
{\bf APPENDIX A:  Analytic Derivation of the $q_i - \overline{V}_i$
Relation Within the Charge Model of Eq. (\ref{charge}).}
\end{center}

Here we derive the $q-V$ relation predicted by the charge
model [Eq. (\ref{charge})], averaging over any fluctuations.
Consider a random $A_{1-x}B_x$ alloy at $x=1/2$
with nearest neighbor
coordination $Z_1$ and an $A$ atom at a central site, denoted
by $A(0)$. (There is no loss of 
generality by choosing this atom to be $A$.)
The charge on $A(0)$ has the distribution
$q_M = -2M\lambda$ ($M=0,Z_1$) with the probability
\renewcommand\arraystretch{0.65}
\begin{equation}
\rho_M = \frac{1}{2^{Z_1}}
\left( \begin{array}{c} Z_1\\M \end{array} \right)
\end{equation}
Therefore, the energy of the random alloy is
\begin{equation}
\langle E_M \rangle = 
\frac{1}{2} \sum_{M=0}^{Z_1} \rho_M q_M 
\sum_m \frac{1}{R_m} q_m(M)
\end{equation}
where $q_m(M)$ is the sum of charge on the $m$th shell
surrounding $A(0)$ under the constraint that there are
$M$ $B(1)$ atoms on the nearest neighbor shell.
$R_m$ is the distance of the $m$th shell atom 
from $A(0)$.  $\langle E_M \rangle$ can also be
written as
\begin{equation}
\langle E_M \rangle = 
\frac{1}{2} \sum_{M=0}^{Z_1} \rho_M q_M 
\overline{V}_M
\end{equation}
where $\overline{V}_M$ is the Coulomb shift on the central site,
averaged over all configurations where there
are $M$ $B(1)$ atoms on the nearest neighbor shell.
Thus, we need to determine $\overline{V}_M$ as a function
of $q_M = -2M\lambda$ where
\begin{equation}
\label{app.shift}
\overline{V}_M = 
\sum_m \frac{1}{R_m} q_m(M).
\end{equation}
In order to compute $\overline{V}_M$, we first need to
compute $q_m(M)$.

{\em First Shell:}
For the nearest-neighbor shell, $m$=1,
\begin{equation}
\label{app.charge.1}
q_1(M) = Mq_1^B + (Z_1-M)q_1^A
\end{equation}
For the $Z_1$ nearest neighbors of an atom in this first shell,
one is $A(0)$, $K_1$ are also nearest neighbors of $A(0)$,
and $\tilde{Z} = Z_1 - K_1 - 1$ are remaining.
For each $A(1)$, the probability that it has
$n$ $B$ neighbors (i.e., with charge $-2n\lambda$)
$l$ of them come from atoms which are not neighbors
of $A(0)$ is
\begin{equation}
\label{app.prob.a.1}
\rho_{n,l}^{A(1)} = 
\left( \begin{array}{c} \tilde{Z}\\l \end{array} \right)
\frac{1}{2^{Z_1-K_1-1}} \;
\frac{
\left( \begin{array}{c} M\\K \end{array} \right)
\left( \begin{array}{c} Z_1-1-M\\K_1-K \end{array} \right)
}{
\left( \begin{array}{c} Z_1-1\\K_1 \end{array} \right)
}
\end{equation}
where $K = n-l$ and
the following inequalities must be satisfied
\begin{eqnarray}
\label{app.prob.a.const.1}
&0& \le n \le Z_1-1 \nonumber \\
&0& \le l \le n \; ; \; l \le \tilde{Z} \nonumber \\
&0& \le n-l \le K_1 \nonumber \\
&n&-l \le M \nonumber \\
&K&_1-(n-1) \le Z_1-1-M
\end{eqnarray}
Similarly, for each $B(1)$, the probability that
it has $n$ $A$ neighbors (i.e., with charge $2n\lambda$),
$l$ of them which are not neighbors of $A(0)$ is:
\begin{equation}
\label{app.prob.b.1}
\rho_{n,l}^{B(1)} =
\left( \begin{array}{c} \tilde{Z}\\l \end{array} \right)
\frac{1}{2^{Z_1-K_1-1}} \;
\frac{
\left( \begin{array}{c} Z_1-M\\K \end{array} \right)
\left( \begin{array}{c} M-1\\K_1-K \end{array} \right)
}{
\left( \begin{array}{c} Z_1-1\\K_1 \end{array} \right)
}
\end{equation}
where $K = n-l-1$ and
the following inequalities must be satisfied
\begin{eqnarray}
\label{app.prob.b.const.1}
&1& \le n \le Z_1 \nonumber \\
&0& \le l \le n-1 \; ; \; l \le \tilde{Z} \nonumber \\
&0& \le n-1-l \le K_1 \nonumber \\
&n&-1-l \le Z_1-M \nonumber \\
&K&_1-(n-1-l) \le M-1
\end{eqnarray}
Combining Eqs. (\ref{app.charge.1}-\ref{app.prob.b.const.1}), we have
\begin{eqnarray}
q_1(M) &=& (Z_1-M)\sum_{n=0}^{Z_1-1} -2n\lambda
\sum_{l=0}^n \rho_{n,l}^{A(1)} \nonumber \\
&+& M\sum_{n=1}^{Z_1} 2n\lambda
\sum_{l=0}^{n-1} \rho_{n,l}^{B(1)}
\end{eqnarray}
where $\rho_{n,l}^{A(1)}$ and $\rho_{n,l}^{B(1)}$ are
subject to the constraints (\ref{app.prob.a.const.1}) and
(\ref{app.prob.b.const.1}).

{\em More Distant Neighbor Shells:}  For $m>1$,
\begin{equation}
\label{app.charge.m}
q_m(M) = \frac{Z_m}{2} [ q_m^A + q_m^B ]
\end{equation}
Atoms on the $m$th shell have $Z_1$ nearest neighbors,
$K_m$ of them are also nearest neighbors of $A(0)$.
Therefore,
\begin{equation}
\label{app.prob.a.m}
\rho_{n,l}^{A(m)} =
\left( \begin{array}{c} Z_1-K_m\\l \end{array} \right)
\frac{1}{2^{Z_1-K_m}} 
\frac{
\left( \begin{array}{c} M\\K \end{array} \right)
\left( \begin{array}{c} Z_1-M\\K_m-K \end{array} \right)
}{
\left( \begin{array}{c} Z_1\\K_m \end{array} \right)
}
\end{equation}
where $K = n-l$ and
the following inequalities must be satisfied
\begin{eqnarray}
\label{app.prob.a.const.m}
&0& \le n \le Z_1 \nonumber \\
&0& \le l \le n \; ; \; l \le Z_1-K_m \nonumber \\
&0& \le n-l \le K_m \nonumber \\
&n&-l \le M \nonumber \\
&K&_m-(n-l) \le Z_1-M
\end{eqnarray}
and
\begin{equation}
\label{app.prob.b.m}
\rho_{n,l}^{B(m)} =
\left( \begin{array}{c} Z_1-K_m\\l \end{array} \right)
\frac{1}{2^{Z_1-K_m}} \;
\frac{
\left( \begin{array}{c} Z_1-M\\K \end{array} \right)
\left( \begin{array}{c} M\\K_m-K \end{array} \right)
}{
\left( \begin{array}{c} Z_1\\K_m \end{array} \right)
}
\end{equation}
where $K = n-l$, subject to the following
constraints
\begin{eqnarray}
\label{app.prob.b.const.m}
&0& \le n \le Z_1 \nonumber \\
&0& \le l \le n \; ; \; l \le Z_1-K_m \nonumber \\
&0& \le n-l \le K_m \nonumber \\
&n&-l \le Z_1-M \nonumber \\
&K&_m-(n-l) \le M
\end{eqnarray}
Combining Eqs. (\ref{app.charge.m}-\ref{app.prob.b.const.m}), we have
\begin{equation}
q_m(M) = \frac{1}{2}Z_m \sum_{n=0}^{Z_1} 2n\lambda
\sum_{l=0}^n [ \rho_{n,l}^{B(m)} - \rho_{n,l}^{A(m)} ]
\end{equation}
where $\rho_{n,l}^{A(m)}$ and $\rho_{n,l}^{B(m)}$ are
subject to the constraints (\ref{app.prob.a.const.m}) and
(\ref{app.prob.b.const.m}).  Note that $q_m(M) = 0$
for any shell which does not share nearest neighbors
with $A(0)$ (i.e., $K_m=0$).
Using the above derived values of $q_1(M)$ and $q_m(M)$
in Eq. (\ref{app.shift}),
we may determine $\overline{V}_M$ 
as a function of $q_M = -2M\lambda$, and as a function of
$M$ this relation is precisely linear, with
no fluctuations:
\begin{equation}
q_M \propto \overline{V}_M.
\end{equation}

\begin{center}
{\bf APPENDIX B: Structural Information for SQS-16}
\end{center}

The ideal (unrelaxed) fcc-based SQS-16 structure ($A_8B_8$) has 
orthorhombic symmetry and primitive lattice vectors 
\begin{equation}
{\bf a} = (\frac{1}{2},\frac{1}{2},0)a \; ; \; \; \;
{\bf b} = (1,-1,2)a \; ; \; \; \;
{\bf c} = (1,-1,-2)a
\end{equation}
The 16 atomic positions, in Cartesian coordinates,
are
\jot=5pt
\begin{eqnarray}
&&A: \; \;        (0,0,0)a                            \nonumber \\
&&B: \; \;        (\frac{1}{2},0,\frac{1}{2})a        \nonumber \\
&&A: \; \;        (\frac{1}{2},-\frac{1}{2},1)a       \nonumber \\
&&B: \; \;        (1,-\frac{1}{2},\frac{3}{2})a       \nonumber \\
&&A: \; \;        (1,-\frac{1}{2},-\frac{3}{2})a      \nonumber \\
&&A: \; \;        (1,-1,-1)a                          \nonumber \\
&&B: \; \;        (\frac{3}{2},-1,-\frac{1}{2})a      \nonumber \\
&&A: \; \;        (\frac{3}{2},-\frac{3}{2},0)a       \nonumber \\
&&B: \; \;        (\frac{1}{2},-\frac{1}{2},-1)a      \nonumber \\
&&A: \; \;        (1,-\frac{1}{2},-\frac{1}{2})a      \nonumber \\
&&B: \; \;        (1,-1,0)a                           \nonumber \\
&&A: \; \;        (\frac{3}{2},-1,\frac{1}{2})a       \nonumber \\
&&A: \; \;        (\frac{1}{2},0,-\frac{1}{2})a       \nonumber \\
&&B: \; \;        (\frac{1}{2},-\frac{1}{2},0)a       \nonumber \\
&&B: \; \;        (1,-\frac{1}{2},\frac{1}{2})a       \nonumber \\
&&B: \; \;        (1,-1,1)a.
\end{eqnarray}
The SQS-16 structure 
matches the first seven pair 
correlation functions of the random $x=1/2$ alloy exactly.


\begin{references}
\bibitem{note.degen}	One might imagine that the equivalence of
			charges in the dilute limit [Eq. (\ref{dilute})]
			could be broken by allowing different values
			of $\lambda$ for $A$ and $B$ atoms,
			$\lambda_A$ and $\lambda_B$.
			However, global neutrality 
			would then require
			$\lambda_A = \lambda_B$.
\bibitem{note.lapw}	The fact that $\lambda_1$ fit to small-unit-cell data 
			is about 15$\%$ smaller than that fit to 
			large-unit-cell data is merely an artifact of the 
			LAPW charges being about 15$\%$ smaller than the 
			multiple scattering charges.
\end{references}
\end{document}